\documentclass[aps,12pt,notitlepage,final,oneside,onecolumn,nobibnotes,
nofootinbib,superscriptaddress,noshowpacs,centertags,amsmath,amssymb,a4paper]{revtex4-2}

\usepackage{geometry}
\usepackage{mathtools}

\newcommand{\bS}{\bm \Sigma}
\newcommand{\bg}{\bm \gamma}

\newcommand{\gz}{\gamma^{0}}
\newcommand{\gf}{\gamma^{5}}

\RequirePackage{graphicx}

\begin{document}

\title{SPECTRAL REPRESENTATION OF NEUTRINO PROPAGATOR IN MEDIA AND OFF-SHELL  SPIN PROPERTIES}

\author{A.E. Kaloshin}
\email{alexander.e.kaloshin@gmail.com}
\affiliation{Irkutsk State University, K.  Marx str., 1, 664003, Irkutsk, Russia}
\affiliation{Joint Institute for Nuclear Research, 141980, Dubna, Russia}

\author{D.M.Voronin}
\email{dmitry.m.voronin@gmail.com}
\affiliation{INR RAS, Moscow 117312, Russia}







\begin{abstract}
   We construct a spectral representation of neutrino propagator in matter moving with constant velocity,  or in constant homogenious magnetic field.  
In both cases there exists definite 4-axis $z$ of complete polarization, such that  corresponding spin projectos commute with propagator. As a result, 
all eigenvalues of propagator and dispersion laws	are classified accoding to spin projection on this axis.
\end{abstract}

\maketitle

When neutrino propagates through the media (dense matter or magnetic field), it modifies a standard picture of flavour oscillations \cite{Wol78, Mik85,OVV86,Akh88_1,Lob01} and further studying needs to use methods of quantum field theory, see e.g. reviews \cite{Beu03,Giu15,Bil18}. 
\footnotetext{Talk at 19th Lomonosov Conf. on Elem. Part. Physics (MSU, Moscow, August 22-28, 2019)}

Here we discuss a convenient algebraic construction for neutrino propagator in media: so-called spectral representation (SR).   In this form, based on the eigenvalue  problem, a propagator looks as a sum of single poles, accompanied by orthogonal matrix projectors.  This representation was discussed earlier for dressed fermion propagator \cite{KL12} and for matrix propagator with mixing of few fermionic fields \cite{KL16}. In case of media a new aspect is the existence of some fixed off-shell 4-axis of complete polarization. It reduces the eigenvalue problem to the vacuum case, media properties only modify  scalar coefficients in general construction.

To construct SR for inverse propagator S, we need to solve the eigenvalue problem (it's better in matrix form, for eigenprogectors $\Pi$)
\begin{equation}\label{eq:5}
S\Pi_i = \lambda_i \Pi_i , \ \ \ \Pi_{i}\Pi_{k}=\delta_{ik}\Pi_{k} .
\end{equation}
After it, S and propagator G can be represented as  sums
\begin{equation}\label{RS}
S=\sum\lambda_{i} \Pi_{i} \ \ \ \Rightarrow \ \ \  G=S^{-1}=\sum \frac{1}{\lambda_{i}} \Pi_{i} .
\end{equation}	
Poles of propagator are  in fact the zeros of eigenvalues $\lambda_{i}$.

In \cite{KL12} this representation was built for dressed propagator of general form
\begin{equation}\label{cov_m}
S=aI+b\hat{p}+c\gf+d\hat{p}\gf  .
\end{equation}
Useful technical step was to introduce the $\gamma$-matrix basis ${\cal P}_{i}$
\begin{equation}\label{basisP}
{\cal P}_{1}=\Lambda^{+},\ \ {\cal P}_{2}=\Lambda^{-},\ \ {\cal P}_{3}=\Lambda^{+}\gamma^{5},\ \ {\cal P}_{4}=\Lambda^{-}\gamma^{5}, \mbox{where}\  \Lambda^{\pm}=\frac{1}{2}(1\pm\frac{\hat{p}}{W})
\end{equation}
with simple algebra. Here $\Lambda^{\pm}$ are off-shell projectors and $W$ is invariant mass $W=\sqrt{p^2}$.
Basis (\ref{basisP}) can be used for decomposition of S and unknown eigenprojector $\Pi$ and
it allows to obtain simple solution of the eigenvalue problem
\begin{eqnarray}\label{vacSR}
\lambda_{i}=\big(S_{1}+S_{2} \pm\sqrt{(S_{1}-S_{2})^{2} + 4 S_{3}S_{4}}\Big)/2~,\ \ \ \ i=1,2 \nonumber\\
\Pi_{i}=(-1)^{i+1}\Big((S_{2}-\lambda_{i}){\cal P}_{1}+(S_{1}-
\lambda_{i}){\cal P}_{2}- S_{3}{\cal P}_{3}-S_{4}{\cal P}_{4}\Big)/(\lambda_{2}-\lambda_{1}),	
\end{eqnarray}
where $S_k$ are coefficients of decomposition of $S$ in the basis (\ref{basisP}).

When neutrino propagates through a matter, there exist two 4-vectors: momentum of particle $p$  and matter velocity $u$, and the most general $\gamma$-matrix decomposition is 
\begin{eqnarray}\label{formula2}
S(p,u)=G^{-1} =s_{1}I+s_{2}\hat{p}+s_{3}\hat{u}+s_{4}\sigma^{\mu\nu}p_{\mu}u_{\nu}+\nonumber\\+s_{5}i\varepsilon^{\mu\nu\lambda\rho}\sigma^{\mu\nu}u_{\lambda}p_{\rho}+s_{6}\gamma^{5}+s_{7}\hat{p}\gamma^{5}+s_{8}\hat{u}\gamma^{5},
\end{eqnarray}
where $s_{i}$ are scalar function dependent on invariants. 

To solve eigenvalue problem for S, the key moment is related with spin properties.
One can introduce 4-vector  $z^{\mu}$, which is a linear combination of $p$,  $u$ and has properties of fermion polarization vector ($z^{\mu}p_{\mu}=0,   z^{2}=-1$):
\begin{equation}\label{vec_z}
z^{\mu}=b\ (p^{\mu}(up)-u^{\mu}p^{2}),\ \ \ \ \ b=[p^{2}((up)^{2}-p^{2})]^{-1/2} .
\end{equation}
Then one can construct (slightly non-standard) the off-shell spin projectors:
\begin{equation}\label{Sigma}
\Sigma^{\pm}=\frac{1}{2}(1\pm\gamma^{5}\hat{z}\hat{n}),~~~ \Sigma^{\pm}\Sigma^{\pm}=\Sigma^{\pm},~~~\Sigma^{\pm}\Sigma^{\mp}=0,\ \ \ n^{\mu}=p^{\mu}/W .
\end{equation}

One can see \cite{KV19} that $\Sigma^{\pm}$ commute with all $\gamma$-matrices in decomposition  (\ref{formula2}). Multiplying the inverse propagator $S(p,u)$ (\ref{formula2}) by unit matrix
\begin{equation}\label{comp}
S=(\Sigma^{+}(z)+\Sigma^{-}(z))S \equiv S^+ + S^- ,
\end{equation}
one obtains two orthogonal contributions  $S^{\sigma},\ \sigma=\pm 1$.

One more useful property of $\Sigma^{\sigma}$ is that in $S^+, S^-$ terms the $\gamma$-matrix structures may be simplified. Namely: $\gamma$-matrices, which contain the matter velocity $u^{\mu}$ may be transformed to the vacuum set of four matrices:  $I, \hat{p}, \gamma^{5}, \hat{p}\gamma^{5}$, see \cite{KV19}. As a result, matter properties are contained only in scalar coefficients.

It can be seen, that  the eigenvalue problem for inverse propagator (\ref{formula2}) is separated into two different problems for $S^{\sigma}$ and these problems are reduced to vacuum case (\ref{basisP}), (\ref{vacSR}) with modified scalar coefficients. 

The introduced 4-vector $z^{\mu}$ (\ref{vec_z}) plays role of the complete polarization axis and all eigenvalues are classified by the projection of spin onto this axis. In contrast to vacuum, this axis is not arbitrary and axis projection is not conserved, see details in  \cite{KV19}.

In particular case of rest matter, polarization vector $z^\mu$  corresponds to helicity state and spin projection is concerved.
In this case the generalized spin projectors (\ref{Sigma}) are projectors onto the spatial momentum direction
\begin{equation}
\Sigma^{\pm}=\frac{1}{2}\Big(1\pm\bS\frac{\bf{p}}{|\bf{p}|}\Big),~~~ \bS=\gamma^{0}\bg\gamma^{5}.
\end{equation}

Thus, for rest matter the well-known fact \cite{Man87,Pan92} is reproduced that neutrino with a definite  dispersion law in matter has a definite helicity.
If  some eigenvalue is vanished $\lambda_{1,2}^\sigma=0$,  we have the known dispersion law
\begin{equation}
E_{1,2}^\sigma=\sigma\alpha\pm\sqrt{(|{\bf{p}}|+\sigma\alpha)^{2}+m^{2}}.
\end{equation}


We found an axis of complete polarization $z^{\mu}$ (\ref{vec_z}) in moving matter.  A similar situation arises when  neutrino propagates in a magnetic field.

Inverse propagator  of neutral fermion with an anomalous magnetic moment $ \mu $ in  magnetic field is
\begin{equation}\label{new_propagator}
S=\hat{p}-m+\mu\bS{\bf{B}},~~~~ \bS=\gz\bg\gf.
\end{equation}

Having electromagnetic field tensor and 4-momentum, we can construct a polarization vector $z^{\mu}$  ($ z^{2}=-1$ and $z_{\mu}p^ {\mu}=0$):
\begin{equation}\label{zB}
z^{\mu}=b\epsilon^{\mu\nu\lambda\rho}F_{\nu\lambda}p_{\rho} \ \ \ \Rightarrow \ \ \ 
z^{\mu}=b(({\bf{B}}{\bf{p}}),p^{0}{\bf{B}}).
\end{equation}
Second expression corresponds to external magnetic field and $b=(p_0^2{\bf B}^2 - ({\bf{B}}{\bf{p}})^2)^{-1/2}$. 
Using this vector, we can construct a standard spin projector:
\begin{equation}\label{sb}
\Sigma^{\pm}=(1 \pm \gamma^{5}\hat{z})/2.
\end{equation}
It is easy to see that the spin projectors commute with the inverse propagator (\ref{new_propagator}): $[S,\Sigma^{\pm}]=0$. 

Further we can apply the same trick that was used in a matter: to obtain two orthogonal terms $S^+$ and $S^-$  (\ref{comp}) and to simplify $\gamma$-matrix structure.
 Instead of (\ref{new_propagator}) we obtain
\begin{equation}\label{s_plus_minus}
S^{\pm}=\Sigma^{\pm}(z)\Big[\hat{p}-m+\frac{\mu}{p^{0}}(\gf({\bf{p}}{\bf{B}}) \pm \gz\frac{1}{b})\Big].
\end{equation}

The inverse propagator in the external field (\ref{new_propagator}), (\ref{s_plus_minus}) is non-covariant, but for algebraic problem this is not so important. At solving of eigenvalue problem with the matrix (\ref{cov_m}), the momentum vector $p^{\mu}$ may be changed by any four numbers. Therefore, if to redefine the vector $p^{\mu}$ in $S^{\pm}$, we can get rid of $\gamma^0$ and use the ready answer (\ref{vacSR}).  
\begin{equation}
p_{\pm}^\mu= (p^{0} \pm \frac{\mu}{bp_{0}}, \ {\bf p} ),\ \ \ \  \ \ 
S^{\pm}=\hat{p}_{\pm}-m+\mu\gf\frac{({\bf{B}}{\bf{p}})}{p_{0}}.
\end{equation} 
Now inverse propagator has 
only $I$, $\hat{p}_{\pm} $ and $\gf$ matrix, and algebraically is similar to the vacuum propagator. Therefore, we can use the formulas (\ref{vacSR}) for eigenvalues and eigenprojectors.

 If eigenvalue is vanishing, we obtain the well-known dispersion law for movement of anomalous magnetic moment in magnetic field \cite{TBKh}
\begin{equation}
E^2 = m^2+ {\bf{p}}^2 + \mu^2 {\bf{B}}^2 \pm 2\mu \sqrt{m^2 {\bf{B}}^2 + {\bf{p}}^2 {\bf{B}}^2_\perp}.
\end{equation}
Here $\pm$ corresponds to different  terms $S^\pm$ in propagator, which are accompanied by spin projectors   $\varSigma^{\pm}$.

So, in constant magnetic field all eigenvalues are classified by the spin projection on the fixed axis $z$ (\ref{zB}). It turns out that, as in the case of moving medium, the projection on this axis, in general, is not a conserved quantity \cite{KV19}.


In summary, we have built the spectral representation (\ref{RS}) of neutrino propagator in a moving matter or in a constant external magnetic field.  The advantage of this representation is that a single term in this sum is related only with one dispersion law for particle in media. 

It turned out that both in matter and in magnetic field there exists the fixed 4-axis of complete polarization $z^\mu$, such that  all eigenvalues of propagator (and, consequently, dispersion laws)	are classified accoding to spin projection onto this axis.  The found spin projectors (\ref{Sigma}), (\ref{sb}) on the axis of complete polarization  play a special role in the eigenvalue problem, simplifying essentially algebraic calculations.  

Note that the found spin projectors  in general case do not commute with Hamiltonian  
$[\varSigma^{\pm}, H]\not=0$. Probably, besides the algebraic simplicity, this axis generate some dynamical spin effects in media.  


\end{document}